\documentclass[pra, reprint, superscriptaddress, floatfix]{revtex4-2}

\usepackage{blindtext}
\usepackage[english]{babel}
\usepackage{amsmath}
\usepackage{amsfonts}
\usepackage{amssymb}
\usepackage{graphicx}
\usepackage{siunitx}
\usepackage{nicefrac}

\begin{document}
\raggedbottom


\title{Spin dynamical decoupling for generating macroscopic superpositions of a free-falling nanodiamond}

\author{B. D. Wood}
\email[]{ben.d.wood@warwick.ac.uk}
\affiliation{Department of Physics, University of Warwick, Coventry, CV4 7AL, United Kingdom}
\author{S. Bose}
\affiliation{Department of Physics and Astronomy, University College London, Gower Street, London, WC1E 6BT, United Kingdom}
\author{G. W. Morley}
\email[]{gavin.morley@warwick.ac.uk}
\affiliation{Department of Physics, University of Warwick, Coventry, CV4 7AL, United Kingdom}

\date{\today}

\begin{abstract}
Levitated nanodiamonds containing negatively charged nitrogen-vacancy centers (${\text{NV}}^{-}$) have been proposed as a platform to generate macroscopic spatial superpositions. Requirements for this include having a long ${\text{NV}}^{-}$ spin coherence time, which necessitates formulating a dynamical decoupling strategy in which the regular spin flips do not cancel the growth of the superposition through the Stern-Gerlach effect in an inhomogeneous magnetic field. Here, we propose a scheme to place a $250 \text{-}\si{\nano\metre}$-diameter diamond in a superposition with spatial separation of over $\SI{250}{\nano\metre}$, while incorporating dynamical decoupling. We achieve this by letting a diamond fall for $\SI{2.4}{\metre}$ through a magnetic structure, including $\SI{1.13}{\metre}$ in an inhomogeneous region generated by magnetic teeth.
\end{abstract}
\maketitle

\section{Introduction}\label{intro}
Quantum mechanics suggests that an object of any size can be placed in a spatial superposition state; however, the most macroscopic superposition observed to date is the matter-wave interference of molecules made of up to $2000$ atoms incident on gratings with period $\SI{266}{\nano\metre}$ \cite{haslinger_2013, eibenberger_2013, fein_2019}. This leads to the question: is there a transition point beyond which a superposition is too macroscopic to exist? To achieve a large macroscopicity a combination of large mass, superposition distance, and time are required \cite{nimmrichter_2013, bassi_2013, frowis_2018}. Creating superpositions with a larger macroscopicity will both increase our knowledge of the domain of validity of quantum mechanics and make progress towards proposed tests of the quantum nature of gravity \cite{bose_2017, marletto_2017, van_de_Kamp_2020, chevalier_2020, kent_2021, kent_garcia_2021, toros_2021, marshman_2021}. One set of proposals for achieving a more macroscopic superposition is based on levitated nanodiamonds, each containing a single nitrogen-vacancy center (${\text{NV}}^{-}$) \cite{yin_2013, scala_2013, wan_2016a, wan_2016b, pedernales_2020}. The core feature of these schemes is that the ${\text{NV}}^{-}$ is put into an electron spin superposition so that an inhomogeneous magnetic field creates a superposition of forces and hence a spatial superposition. The two components of this matter wave then interfere with each other to produce fringes. Here we build on the idea of using a free-falling nanodiamond \cite{wan_2016b}.   

The inhomogeneous magnetic field used to create the superposition will also provide diamagnetic trapping in one dimension \cite{pedernales_2020}. A scheme has been proposed that dynamically decouples the ${\text{NV}}^{-}$ spin, increases the maximum separation distance of spatial superposition, and dynamically decouples the matter wave from decoherence \cite{pedernales_2020}. However, for realistic magnetic fields, the oscillation period is too long to provide effective spin decoupling; therefore, we introduce a structure of magnetic teeth. The proposed scheme does not require the motion of the center of mass of the nanodiamond to be cooled to the quantum ground state \cite{scala_2013, wan_2016b, pedernales_2020}, although motional ground-state cooling has been experimentally demonstrated for levitated nanoparticles \cite{delic_2020, magrini_2021, tebbenjohanns_2021}. Other key experiments in this area include nanodiamonds levitated with optical tweezers \cite{hoang_2016b, neukirch_2015, pettit_2017, neukirch_2013, rahman_2016, frangeskou_2018}, Paul traps \cite{delord_2017b, delord_2017a, delord_2018, kuhlicke_2014, conangla_2018, delord_2020}, and magnetic traps \cite{hsu_2016, obrien_2019}. The rotational motion of levitated nanodiamonds has also been shown to be relevant in theory \cite{ma_2017} and experiment \cite{hoang_2016a}. 

A key requirement of this proposal is that the ${\text{NV}}^{-}$ center contained in the falling nanodiamond must have a long enough spin coherence time, ${T}_{2}$, to remain coherent throughout the drop. Using dynamical decoupling, ${T}_{2}$ times exceeding $\SI{1}{\second}$ have been measured for ${\text{NV}}^{-}$ in nonlevitated bulk diamonds \cite{abobeih_2018, bar-gill_2013}. The longest reported ${T}_{2}$ time in micro- or nanodiamonds is $\SI{708}{\micro\second}$ using isotopically pure ${}^{12}\text{C}$ diamond material that is then etched into pillars of diameters $300$ to $\SI{500}{\nano\metre}$ and lengths $\SI{500}{\nano\metre}$ to $\SI{2}{\micro\metre}$ \cite{andrich_2014}. For natural abundance ${}^{13}\text{C}$ micro- or nanodiamonds, the longest ${T}_{2}$ time reported for particles fabricated using etching techniques is $\SI{210}{\micro\second}$ \cite{trusheim_2014} and by milling $\SI{460}{\micro\second}$ \cite{wood_2021c, knowles_2014}. 

\begin{figure}
	\includegraphics[width=\linewidth]{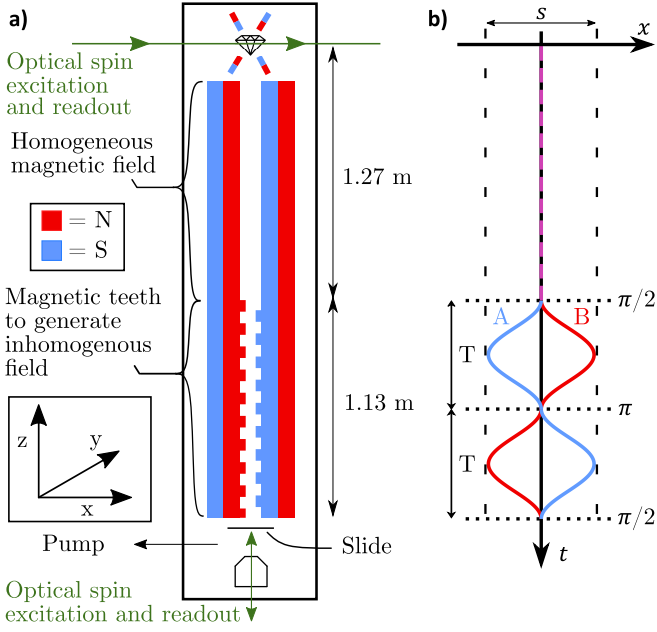}
	\caption{(a) Schematic of the proposed experiment for testing macroscopic superpositions. A nanodiamond should be cryogenically cooled to $\SI{5}{\kelvin}$ on a cold surface and then shaken off with ultrasound to be levitated in a magnetic trap or Paul trap in ultrahigh vacuum. The diamond is then dropped through a magnetic field that includes homogeneous and inhomogeneous regions. The diamond lands on a glass slide held at $\SI{5}{\kelvin}$. Magnets are shown schematically with their north and south poles in red (dark gray) and blue (light gray), respectively. The schematic is not to scale. (b) Schematic of the paths taken by the two superposition components, $A$ in solid blue (solid light gray), $B$ in solid red (solid dark gray), and dashed purple (dashed gray) for presuperposition. $T$ is the period of oscillation of the separation and $s$ the maximum separation reached. \label{schematic}}
\end{figure}

\section{Proposed Scheme}\label{scheme}

\begin{figure*}
	\includegraphics[width=0.728\linewidth]{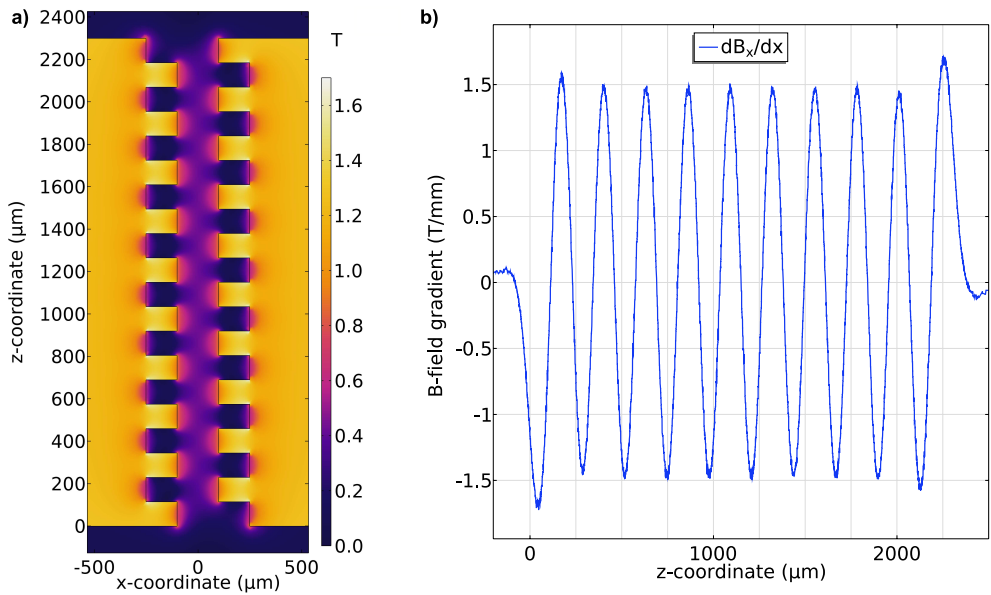}
	\caption{Finite element simulations, using COMSOL multiphysics software, of a shorter length of the proposed magnetic tooth structure to produce an alternating inhomogeneous magnetic field, generating a spatial superposition in the $x$ direction. Neodymium magnets are considered, both magnetized in the $+x$ direction. (a) A $y=0$ plane plot of the magnitude of the magnetic field. The flux concentrates in the teeth, producing large field gradients across the gap as the opposing teeth are offset from each other by the tooth width. (b) The gradient of the $x$ component of the magnetic field in $x$ direction, simulated along the $z$ axis of the magnetic geometry. The noise present is due to the finite size of mesh elements in COMSOL. The gradient of the $y$ and $z$ components of the magnetic field are plotted in Fig. \ref{z_ax} in Appendix \ref{a1}. \label{COMSOL}}
\end{figure*}

We now go on to propose a method for generating a macroscopic spatial superposition of nanodiamonds, shown schematically in Fig. \ref{schematic}. In free fall, the Hamiltonian that governs the dynamics of the nanodiamond in the $x$ direction is given by \cite{wan_2016b, pedernales_2020}

\begin{equation}\label{eq:1}
\begin{split}
H = &\frac{{{\hat{p}}_{x}}^{2}}{2m} + {g}_{\parallel} {\mu}_{\text{B}} (\pm{B}^{\prime} \hat{x}+{B}_{0}) {\hat{S}}_{{z}^{\prime}} + \frac{\lvert \chi\rvert V}{2 {\mu}_{0}} {(\pm{B}^{\prime} \hat{x}+{B}_{0})}^{2}\\
 &+ mg \sin(\phi) \hat{x} + \hbar D {{\hat{S}}_{{z}^{\prime}}}^{2}, 
\end{split}
\end{equation}
where $\phi$ is the tilt between the z axis defined by the magnet geometry and the vertical defined by gravity. To estimate the maximum spatial separation, the external magnetic field is assumed to have the form $ \mathbf{B}(\hat{\mathbf{r}}) = (\pm{B}^{\prime} \hat{x}+{B}_{0}) {\mathbf{e}}_{x}$, where $\pm{B}^{\prime}$ is the average magnitude of the magnetic-field gradient, ${B}^{\prime} = \SI{940}{\tesla\per\metre}$, changing direction with each magnetic tooth crossing, and ${B}_{0} = \SI{420}{\milli\tesla}$ is the bias magnetic field at $x=0$; see Fig. \ref{COMSOL}. The ${\text{NV}}^{-}$ axis, ${z}^{\prime}$, is assumed to be aligned with the $x$ axis.

The first term of Eq. (\ref{eq:1}) describes the kinetic energy of the nanodiamond, the second is due to the energy of the ${\text{NV}}^{-}$ electron spin magnetic moment in the external magnetic field, the third includes the diamagnetic effect of the diamond material in the dynamics, the fourth term describes the relative gravitational potential energy for different $x$ positions in the presence of tilt, and the final term is due to the zero-field splitting of the ${\text{NV}}^{-}$ center energy levels. ${\hat{p}}_{x}$ and $\hat{x}$ are the momentum and position operators along the $x$ axis, respectively. $m = \SI{2.9e-17}{\kilo\gram}$, $V = \SI{8.2e-21}{\metre\cubed}$, and $\chi = \SI{-2.2e-5}{}$ are the mass, volume, and volume magnetic susceptibility of the nanodiamond. The proposed mass and volume correspond to a nanodiamond containing around $1.5$ billion atoms with a diameter of $\SI{250}{\nano\metre}$. ${g}_{\parallel}= 2.0029$ \cite{felton_2009} and ${\hat{S}}_{{z}^{\prime}}$ are the $g$ factor and spin component operator for an ${\text{NV}}^{-}$ with its axis assumed to be aligned with the external magnetic field. $D = \SI{2.87}{\giga\hertz}$ is the zero-field splitting of ${\text{NV}}^{-}$ \cite{doherty_2013}. ${\mu}_{\text{B}}$, ${\mu}_{0}$, and $g$ are the Bohr magneton, vacuum permeability, and acceleration due to gravity, respectively. A detailed description of the dynamics is provided in Appendixes \ref{a1} and \ref{a2}.

The experimental protocol would run as described in the following 11 steps.
(1) Trap a nanodiamond and measure the diamond mass by fitting the power-spectral density \cite{chang_2010, rahman_2018}. 
(2) Collect an optical fluorescence spectrum, to confirm the presence of ${\text{NV}}^{-}$. 
(3) Perform a Hanbury Brown-Twiss (HBT) measurement to confirm the trapped nanodiamond contains a single ${\text{NV}}^{-}$.
(4) Measure the ODMR spectrum to check the orientation of the ${\text{NV}}^{-}$. Align the ${\text{NV}}^{-}$ axis to the $x$ axis using a magnetic \cite{ma_2017, delord_2020} or electric field \cite{pedernales_2019}. Electrically neutralize the diamond with a radioactive source and/or UV light \cite{hsu_2016, ranjit_2016, moore_2014}. Steps (1) to (4) would take around five minutes for each nanodiamond. 
(5) Optically polarize the ${\text{NV}}^{-}$  spin to $|0\rangle$, apply a $\pi$ pulse to transfer to $|-1\rangle$, and drop the diamond. The first $\SI{1.27}{\metre}$ of the drop occurs in a homogeneous magnetic field aligned along the $+x$ direction. This keeps the ${\text{NV}}^{-}$ aligned along the $x$ axis and allows the nanodiamond to build up speed so that the magnetic teeth in the second half of the drop are crossed at a frequency that can be used for dynamical decoupling.  
(6) After dropping $\SI{1.27}{\metre}$, apply a microwave $\pi/2$ pulse to change the spin state from $|0\rangle$ to $ (|0\rangle + |-1\rangle)/ \sqrt{2}$.  The superposition distance could be doubled by using the state $(|-1\rangle + |1\rangle)/\sqrt{2}$ instead \cite{fang_2013, mamin_2014}, but this would increase the number of pulses required for the dynamical decoupling, which would make the decoupling less efficient, reducing the ${T}_{2}$.
(7) An alternating inhomogeneous magnetic field is present throughout the final $\SI{1.13}{\metre}$ of the drop, which lasts $\SI{190}{\milli\second}$. The teeth mean that the magnetic-field inhomogeneity alternates between $\SI{\pm1.45}{\tesla\per\milli\metre}$, as shown in the simulations in Fig. \ref{COMSOL}(b). After going into the initial spin superposition, the diamond first experiences a strong magnetic-field gradient in the $+x$ direction, which provides a superposition of forces and creates a small spatial superposition. As the diamond falls into a region of magnetic-field gradient in the $-x$ direction a microwave $\pi$ pulse is applied. This flips the spin states within the superposition, so that the spatial superposition distance will continue to increase. The repeated $\pi$ pulses also refocus the spin decoherence from the spin bath of the diamond. The phases of the $\pi$ pulses are chosen to provide dynamical decoupling, such as a modified XY8 sequence \cite{wang_2012}. Unlike the standard XY8 sequence, the time between pulses decreases throughout the drop as the nanodiamond continues to accelerate. 
(8) Recombination of superposition components is due to diamagnetic forces \cite{scala_2013, pedernales_2020}, which is an effect of the harmonic form of the trapping potential.  After $\SI{23.75}{\milli\second}$, the superposition distance generates enough of a differential in the diamagnetic acceleration experienced by the two components of the superposition that there is no relative acceleration. Beyond this superposition distance the components accelerate towards each other, with a maximum separation reached after $\SI{47.5}{\milli\second}$. The spin and diamagnetic forces will recombine the superposition components after $\SI{95}{\milli\second}$. At this point an extra $\pi$ pulse is applied, for motional dynamical decoupling \cite{pedernales_2020}, and the drop continues for a further $\SI{95}{\milli\second}$ until a second recombination so that, in the second half, each spatial superposition component follows the path the other took in the first half, as shown in Fig. \ref{schematic}(b). After $\SI{190}{\milli\second}$ a $\pi/2$ pulse is applied to end the interferometric scheme and the nanodiamond is caught on a glass slide which is held at $\SI{5}{\kelvin}$.
(9) Optically readout the spin state of the ${\text{NV}}^{-}$.
(10) Move the slide slightly and repeat from step (1).
(11) Measure the ${T}_{2}$ of each of the ${\text{NV}}^{-}$ in different nanodiamonds in a row on the glass slide. Diamonds with ${T}_{2}$ times that are short compared with the drop time would not be included in the data analysis of the spatial superposition.

The maximum superposition distance created can be estimated by $s= (2{g}_{\parallel}{\mu}_{\text{B}}{\mu}_{0})/(V \lvert \chi\rvert {B}^{\prime})$ \cite{pedernales_2020}. For the parameters proposed in Eq. (\ref{eq:1}), $s=\SI{276}{\nano\metre}$. The oscillation frequency of the superposition distance due to diamagnetic forces is $\omega = \sqrt{\lvert \chi\rvert /\rho{\mu}_{0}}{B}^{\prime} = (2\pi)\SI{10.56}{\hertz}$ \cite{pedernales_2020}, where $\rho = \SI{3510}{\kilo\gram\per\metre\cubed}$ is the density of diamond. 

\section{Discussion}\label{disc}

Optical traps raise the internal temperature of impure nanodiamonds in vacuum \cite{rahman_2016}, and even high-purity nanodiamonds would absorb enough of the trapping light to raise the internal temperature by hundreds of kelvin at gas pressures of $\SI{E-6}{\milli\bar}$ \cite{frangeskou_2018}. High internal temperatures of the trapped nanodiamond will limit ${\text{NV}}^{-}$ $T_{2}$ times \cite{bar-gill_2013} and blackbody radiation can collapse the spatial superposition \cite{romero-isart_2011}, even if the motion of the center of mass is cooled by feedback cooling. Therefore, a magnetic trap \cite{hsu_2016, obrien_2019} or Paul trap \cite{delord_2017b, delord_2017a, delord_2018, kuhlicke_2014, conangla_2018} should be used to reduce internal heating. Dropping the diamond from a magnetic trap could be achieved by applying an electric field to pull the diamond out of the bottom of the trap \cite{hsu_2016}. This would use the electric dipole induced in the diamond. Dropping the diamond from a Paul trap may be possible by neutralizing the particle.  

Before the drop, the rotational vibrations would have a temperature of \SI{5}{\kelvin}. There is no need to reach the ground state \cite{scala_2013, wan_2016b} but the diamond should maintain its orientation during the drop. This is needed to avoid torques \cite{pedernales_2019}, to ensure a predictable ODMR spectrum for the microwave pulses, and to maximize the force experienced \cite{pedernales_2019}. As the room-temperature ${\text{NV}}^{-}$ ${T}_{2}$ is on the order of $\SI{}{\milli\second}$, the homogeneous linewidth should be below $\SI{30}{\nano\tesla}$, so the $g$ factor should be defined to one part in ${10}^{6}$; however, it depends on the orientation of the external magnetic field relative to the ${\text{NV}}^{-}$ axis \cite{felton_2009}. One way to make the $g$ factor repeatable to one part in ${10}^{6}$ would be to hold the orientation of the ${\text{NV}}^{-}$ to within $\pm 4$ degrees of the external field. The diamond remains in a magnetic field aligned along the $x$ axis throughout the drop; this will keep the ${\text{NV}}^{-}$ axis oscillating around the $x$ direction \cite{ma_2017, delord_2020}. To avoid precession \cite{jackson_kimball_2016} feedback cooling \cite{hsu_2016, gieseler_2012, delord_2020} of the rotation of the diamond is required before the drop. Light should not be used to provide this feedback because of its heating effects, but dielectric electrical detection and feedback would avoid this heating \cite{hsu_2016, goldwater_2019}. 

As well as cooling rotational vibrations prior to the drop, stray surface spins on the nanodiamond should be polarized using a large superconducting magnet, generating approximately $\SI{8}{\tesla}$. Spin bath polarization by magnetic fields of this magnitude have been shown to extend the ${T}_{1}$ and ${T}_{2}$ times of both the ${\text{NV}}^{-}$ spin and the spin bath by suppressing energy-conserving flip-flop transitions \cite{takahashi_2008}. Alongside extending ${\text{NV}}^{-}$ spin coherence time, this will reduce noise due to spin bath transitions both during each drop and from drop to drop. ODMR at high magnetic fields has also been previously demonstrated in ${\text{NV}}^{-}$ centers \cite{aslam_2015, stepanov_2015}. This strong magnetic field will also aid diamagnetic trapping. Assuming the extra field is uniform across the trapping region, the magnet will induce a greater magnetization in the nanodiamond, without affecting the shape of the trap. Even if a field of only a few tesla could be generated, the benefits would still be present to a lesser extent.  

The magnitude of the inhomogeneous magnetic field must be extremely stable not only between drops, to build an interference pattern without washing out fringes, but also during each drop to maintain a coherent state through to the recombination of the two spatial superposition components. The problem of achieving coherent recombination via magnetic fields has been analyzed in the "Humpty-Dumpty" effect \cite{englert_1988, schwinger_1988, scully_1989}. Despite the extreme accuracy required, coherent recombination has been demonstrated experimentally for a Bose-Einstein condensate \cite{margalit_2021, keil_2021}. Bose-Einstein condensates have also been proposed as another approach to probe the quantum nature of gravity \cite{howl_2019}. 
  
The proposed magnetic tooth width in the $z$ direction of $\SI{115}{\micro\metre}$ sets the number of microwave $\pi$ pulses used for dynamical decoupling during the inhomogeneous field region of the drop at around $9800$. The ${\text{NV}}^{-}$ coherence times of $\SI{600}{\milli\second}$ and $\SI{1.58}{\second}$ were achieved with $8192$ and $10240$ pulses, respectively \cite{bar-gill_2013, abobeih_2018}. The magnetic field at the location of the diamond in the inhomogeneous region is, from Fig. \ref{COMSOL}(a), approximately $\SI{420}{\milli\tesla}$; therefore, Majorana spin flips of the unwanted electronic and nuclear spin impurities are avoided \cite{marshman_2021}. 

Continuous microwave excitation could also be used between discrete $\pi$ pulses to implement spin locking of the ${\text{NV}}^{-}$. Here microwaves drive rotations of the spin state about the axis along which it is aligned in the Bloch sphere \cite{levitt_2001, schweiger_jeschke_2001, naydenov_2011}. This alternative form of dynamical decoupling would allow fewer teeth to be used, as coherence is extended without flipping the spin state; however, the magnetic teeth cannot be removed entirely. Without the magnetic-field gradient alternating in direction a large diamagnetic oscillation can move the nanodiamond into a significantly different magnetic-field region, or even crash the nanodiamond into the magnets. The number of magnetic teeth could be reduced from $9800$ to a few hundred to avoid large diamagnetic oscillation, if spin locking adequately maintains coherence between $\pi$ pulses. Spin locking could also reduce the length of the drop prior to the magnetic teeth, as $\pi$ pulses are not needed as frequently to maintain coherence so the required falling speed of the diamond prior to magnetic teeth is reduced.   

By ensuring that the two spatial superposition components follow the same, time-inverted, path across the first and second oscillation, as shown in Fig. \ref{schematic}(b), the accumulation of relative phase between the two superposition components due to static potentials that have the same dependence on $x$ in the first and second oscillation are suppressed \cite{pedernales_2019, pedernales_2020}. The induced gravitational phase due to a tilt of the magnets from the vertical defined by gravity is partially suppressed \cite{scala_2013, wan_2016b, pedernales_2019}. The uncancelled phase is due to the change in the gravitational acceleration as the nanodiamond gets closer to the center of mass of the Earth during the drop. The fringe width of the interference pattern produced is sampled by sweeping $\phi$ by approximately $\pm 500 ~\si{\micro\radian}$ around $\phi=0$. Active stabilization of an optical table to less than $\SI{350}{\nano\radian}$ of tilt noise has been demonstrated \cite{lewandowski_2020}. The fringe width could be reduced by increasing the magnet separation in the second oscillation. The reduction in magnetic-field gradient will increase both the superposition distance and the phase accumulated in the second oscillation.  This mechanism could be used to evidence superposition with an effect that couples only to the nanodiamond, not the ${\text{NV}}^{-}$ spin. Other suppressed relative phase contributions include the magnetic moments from non-${\text{NV}}^{-}$ spins in the nanodiamond and the nanodiamond's electric dipole moment \cite{pedernales_2019}. The width and separation of the magnetic teeth are kept constant throughout the drop to maintain a consistent spatial superposition separation in the first and second oscillation. Therefore, as the nanodiamond continues to accelerate under gravity, the nanodiamond covers a greater distance in the $z$ direction and a greater number of teeth are crossed in the second oscillation than in the first.

Larger superposition distances could be reached by increasing the drop time and implementing more motional dynamical decoupling oscillations \cite{pedernales_2020} in addition to the spin decoupling provided by the magnetic teeth and $\pi$ pulses. Alternatively, by inserting homogeneous field regions in the magnetic teeth structure at the point at which the rate of gain of separation of the superposition components is greatest, the superposition could be allowed to evolve freely. This increases the maximum separation reached before inhomogeneous magnetic fields can be used to recombine the superposition.  

The traditional room-temperature spin readout of the ${\text{NV}}^{-}$ has a single-shot signal-to-noise ratio (SNR) of only $0.03$, so would require over ${10}^{5}$ nanodiamonds to be dropped just to get a single data point with $\text{SNR} = 10$ \cite{hopper_2018}. However, by keeping the diamond at $\SI{5}{\kelvin}$, single-shot single-spin readout can be used via photoluminescence excitation. $95\%$ ${\text{NV}}^{-}$ readout fidelity has been demonstrated \cite{abobeih_2018, robledo_2011}.

Decoherence of the matter wave can come from gas atoms and from blackbody radiation. It has been shown that $\SI{5}{\kelvin}$ is cold enough that blackbody radiation will not cause wave-function collapse even up to much larger superposition distances \cite{romero-isart_2011}. Pressures of below $\SI{e-13}{\milli\bar}$ would be needed to avoid decoherence from gas atoms as a single collision would cause collapse for a superposition distance of $\SI{276}{\nano\metre}$ \cite{romero-isart_2011}.

A generic obstacle to any such large superpositions as are described in this proposal is that the phases developed are so large that they can be pseudo-random unless the environment is controlled precisely. To avoid this problem, initial runs would use a much shorter superposition time and distance, with the superposition recombining due to extra $\pi$ pulses \cite{wan_2016b} rather than solely diamagnetic forces. By slowly increasing the superposition distance and time, accumulated phases could be identified and controlled before they become so large that they are pseudorandom. This additional phase is generated by sources such as external electric and magnetic fields, and tilt instability. These phases could be actively canceled. For example, by monitoring external electric fields and then using electrodes to output a canceling field in real time, additional phase accumulation is reduced. The additional phase could also be canceled by monitoring the external sources, and then subtracting their effects from the measured spin state data.  

The timing precision required is on the order of $\SI{1}{\nano\second}$, which is reasonable as microwave pulses are routinely applied with nanosecond precision in ${\text{NV}}^{-}$ experiments. Thermal expansion of the magnets can reduce the maximum spatial superposition distance due to mistiming of the $\pi$ pulses. To suppress induced phases the magnets need to be held at a temperature stable to better than $\SI{1}{\milli\kelvin}$, or the length of the drop monitored to better than $\SI{10}{\nano\metre}$ precision and the pulse timings adjusted accordingly.

The mechanism used to drop the nanodiamond from the initial trap may not be consistent enough to rely on for the required submicrosecond timing precision required. Instead, a timing gate system in the first half of the free fall (in the homogeneous field region prior to the teeth) could be used to synchronize the $\pi$ pulse timing. Low-power photon beams across the drop region could be used to detect the passage of the nanodiamond in free fall, before the superposition scheme begins. Once the timings of the nanodiamond passing the timings gates are known, the $\pi$ pulses can be timed to the tooth crossings.

The spin must be read out at the time of the second recombination as otherwise the superposition components will spatially separate again. Therefore, a slide should be placed at the vertical position of the second recombination of the spatial states to catch the nanodiamond without collapsing the spin state superposition. The spin can then be read out from the stationary nanodiamond in a confocal microscope setup. The slide must be positioned vertically with enough precision such that the nanodiamond superposition state colliding with it does not provide the required which path information to collapse the spin superposition. That is, the collision must occur when the spatial state and the spin state of the ${\text{NV}}^{-}$ are separable, as discussed in Appendix \ref{a2}. 

\section{Conclusion\label{conc}}
We have proposed a scheme to place a $\SI{250}{\nano\metre}$ diameter diamond into a spatial superposition with a separation of over $\SI{250}{\nano\metre}$. This would probe the macroscopic limits of quantum mechanics. The scheme incorporates spin dynamical decoupling pulses, which is crucial to enable the spin coherence times required to generate such a macroscopic superposition. Spin dynamical decoupling is possible due to a proposed structure of magnetic teeth. The paths of each superposition component in the first and second half of the drop are chosen such that the only gravitational phase accumulated is due to the change in the gravitational field down the drop. This relaxes the requirements on tilt stability. 

It has previously been proposed that two macroscopic superpositions interacting gravitationally could be used to probe the quantum nature of gravity \cite{bose_2017, marletto_2017}. The magnetic structure we propose here could be scaled up to implement the test of quantum gravity in the arrangement labeled "symmetric" in Ref. \cite{nguyen_2020} or "parallel" in Ref. \cite{tilly_2021}. By increasing the depth of the magnets in the $y$ direction, and dropping two diamonds offset in $y$, two identical superpositions could be generated simultaneously. This is advantageous, as generating two superpositions with separate magnets and control infrastructure in close proximity would be extremely challenging.  

\begin{acknowledgments}
S.B. acknowledges EPSRC Grants No. EP/N031105/1 and No. EP/S000267/1. This work is supported by the UK National Quantum Technologies Programme through the NQIT Hub (Networked Quantum Information Technologies), the Quantum Computing and Simulation (QCS) Hub, and the Quantum Technology Hub for Sensors and Metrology with funding from UKRI EPSRC Grants No. EP/M013243/1, No. EP/T001062/1, and No. EP/M013294/1, respectively. G.W.M. is supported by the Royal Society. 
\end{acknowledgments}

\appendix
\section{Nanodiamond Motion}\label{a1}

\begin{figure*}
	\includegraphics[width=0.67\linewidth]{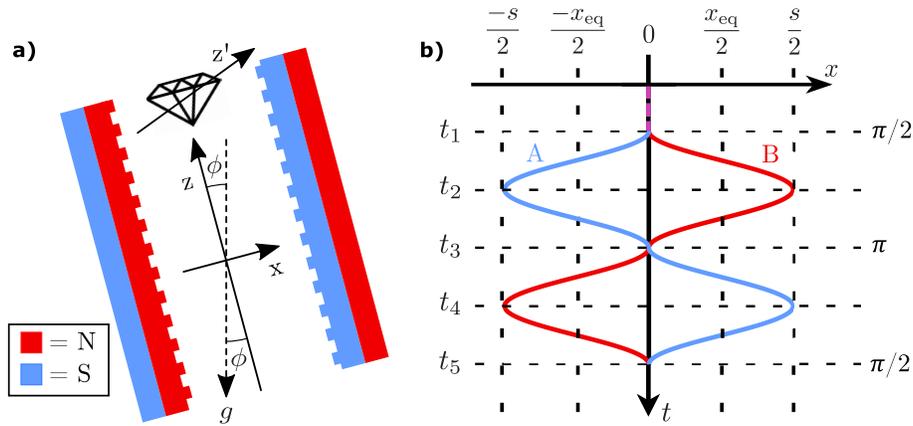}
	\caption{(a) Schematic showing the coordinate axes for the spatial superposition generation scheme. $x$, $y$ (not shown), and $z$ axes are defined in the reference frame of the magnets. Any tilt to the magnets creates an angle $\phi$ with the vertical defined by gravity. The orientation of the axis of the ${\text{NV}}^{-}$ center in the falling nanodiamond is given by ${z}^{\prime}$. Magnets are shown schematically with their north and south poles in red (dark gray) and blue (light gray) respectively. The magnitude of $\phi$ is exaggerated. (b) An annotated and scaled copy of Fig. \ref{schematic}(b), depicting the paths taken by the two spatial superposition components, one in solid red (solid dark gray), the other in solid blue (solid light gray), and dashed purple (dashed gray) for presuperposition. $s$ is the maximum separation reached. Key microwave pulses are indicated, but the spin dynamical decoupling spin flips that coincide with crossing magnetic teeth are not shown for clarity. $\Delta{x}_{\text{eq}}$ indicates the separation of equilibrium positions about which the paths oscillate. \label{geom}}
\end{figure*}

Let us restate the geometry and Hamiltonian considered for the dynamics of the falling nanodiamond in Sec. \ref{scheme}. In the reference frame of the magnets the $x$ axis is the magnetization axis and the z axis runs parallel to the gap between the magnets, as shown in Fig. \ref{geom}(a). A tilt in the magnetic structure will create an angle $\phi$ between the $z$ axis and the vertical defined by gravity.

\begin{equation}\label{eq:s1}
\begin{split}
H = &\frac{{{\hat{p}}_{x}}^{2}}{2m} + {g}_{\parallel} {\mu}_{\text{B}} (\pm{B}^{\prime} \hat{x}+{B}_{0}) {\hat{S}}_{{z}^{\prime}} + \frac{\lvert \chi\rvert V}{2 {\mu}_{0}} {(\pm{B}^{\prime} \hat{x}+{B}_{0})}^{2}\\
&+ mg \sin(\phi) \hat{x} + \hbar D {{\hat{S}}_{{z}^{\prime}}}^{2} .
\end{split}
\end{equation}

Only dynamics of the nanodiamond in the $x$ axis are considered by the Hamiltonian in Eq. (\ref{eq:s1}) \cite{wan_2016b, pedernales_2020}, with the free fall occurring due to gravity assumed to be independent for small $\phi$. The external magnetic field is assumed to have the form $ \mathbf{B}(\hat{\mathbf{r}}) = (\pm{B}^{\prime} \hat{x}+{B}_{0}) {\mathbf{e}}_{x}$, where $\pm{B}^{\prime}$ is the average magnitude of the magnetic-field gradient, ${B}^{\prime}$, changing direction with each magnetic tooth crossing, and ${B}_{0}$ is the bias magnetic field at $x=0$. The ${\text{NV}}^{-}$ axis, ${z}^{\prime}$, is assumed to be aligned with the $x$ axis.

\begin{figure*}
	\includegraphics[width=0.5\linewidth]{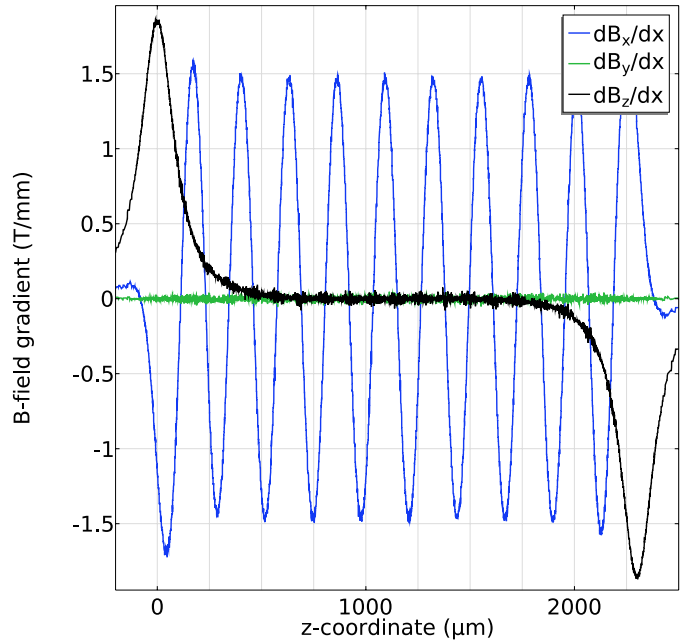}
	\caption{Gradient of the $y$, in green (light gray), and $z$, in black (black), components of the magnetic field as well as the $x$ component, in blue (dark gray) displayed in Fig. \ref{COMSOL}. Away from edge effects, along the $z$ axis of the magnet geometry, the only field gradient in the $x$ direction is from the $x$ component of the magnetic field. As is the case in Fig. \ref{COMSOL}, the noise present is due to the finite size of mesh elements in COMSOL. \label{z_ax}}
\end{figure*}

Before discussing the spatial superposition generation scheme presented in Sec. \ref{scheme}, first let us consider the dynamics in a static magnetic-field gradient, without teeth, tilt, or bias field. In this case, the equilibrium $x$ position of the $|0\rangle$ and $|-1\rangle$ spin states of the ${\text{NV}}^{-}$ center are separated by $\Delta {x}_{\text{eq}}=({g}_{\parallel}{\mu}_{\text{B}}{\mu}_{0})/(V \lvert \chi\rvert {B}^{\prime})$ \cite{pedernales_2020}. 

However, the magnetic tooth structure is required to incorporate spin dynamical decoupling of a useful frequency into the schemes previously proposed \cite{wan_2016b, pedernales_2020}. The diamagnetic effect of the oscillating magnetic-field gradient direction can be separated into two components. First, a force independent of $x$ position acts on the nanodiamond, $F \propto \mp {B}^{\prime}{B}_{0}$. This causes the nanodiamond position to oscillate around $x=0$ as it falls. This occurs due to the nonzero magnetic field at $x=0$ and acts independently of spin state; thus each spatial superposition component feels the same force throughout the drop. For this reason in further discussions of the relative dynamics this effect is ignored. Secondly, the nanodiamond experiences a restoring diamagnetic force, $F \propto -{(\pm {B}^{\prime})}^{2}\hat{x}$. This effect only depends on the magnetic-field gradient, not the bias. Furthermore, due to the spin flips and swapping of paths $A$ and $B$, both superposition components contain an ${\text{NV}}^{-}$ in the spin $|0\rangle$ and $|-1\rangle$ states for the same amount of time; therefore, the constant bias field term in the spin magnetic moment term can be canceled. Hence the new considered Hamiltonian is given by  
  
\begin{equation}\label{eq:s2}
\begin{split}
H = &\frac{{{\hat{p}}_{x}}^{2}}{2m} + {g}_{\parallel} {\mu}_{\text{B}} (\pm{B}^{\prime} \hat{x}) {\hat{S}}_{{z}^{\prime}} + \frac{\lvert \chi\rvert V}{2 {\mu}_{0}} {(\pm{B}^{\prime} \hat{x})}^{2} + mg \sin(\phi) \hat{x}\\
&+ \hbar D {{\hat{S}}_{{z}^{\prime}}}^{2}.
\end{split}
\end{equation}

In the absence of tilt this is the same as the Hamiltonian presented in Ref. \cite{pedernales_2020}, except for the $\pm{B}^{\prime}$ direction changes. However, the alternating direction of the magnetic-field gradient generated by the magnetic teeth combined with the spin flipping $\pi$ pulses timed to coincide with the nanodiamond crossing the change in field gradient direction mean that the spatial superposition components experience the same relative forces as if the spin was not flipping, and the field gradient was not changing direction. Therefore, the final Hamiltonian can be written: 

\begin{equation}\label{eq:s3}
\begin{split}
H = &\frac{{{\hat{p}}_{x}}^{2}}{2m} + {g}_{\parallel} {\mu}_{\text{B}} {B}^{\prime} \hat{x} {\hat{S}}_{{z}^{\prime}} + \frac{\lvert \chi\rvert V}{2 {\mu}_{0}} {({B}^{\prime} \hat{x})}^{2} + mg \sin(\phi) \hat{x}\\
&+ \hbar D {{\hat{S}}_{{z}^{\prime}}}^{2}. 
\end{split}
\end{equation}

Here the relative position dynamics of the two superposition components can be explained in an effective scheme, where there are no spin flips, other than the microwave pulses labeled in Fig. \ref{geom}(b), and the magnetic-field gradient has a constant magnitude and direction with zero bias field. However, in the absence of tilt, this is exactly the scheme described in Ref. \cite{pedernales_2020}. Therefore, the spatial separation of the two superposition components can be described by two harmonic oscillators with $\Delta {x}_{\text{eq}}=({g}_{\parallel}{\mu}_{\text{B}}{\mu}_{0})/(V \lvert \chi\rvert {B}^{\prime})$ and $\omega = \sqrt{\lvert \chi\rvert /\rho{\mu}_{0}}{B}^{\prime}$ \cite{pedernales_2020}. 

Although the relative forces on each superposition component are identical, there is a difference between the scheme presented here and that presented in Ref. \cite{pedernales_2020}. Due to the combined spin flipping and magnetic-field gradient direction change, in the first oscillation shown in Fig. \ref{geom}(b), whenever path $A$ contains spin $|-1 \rangle$ it "sees" an equilibrium position of $-\Delta{x}_{\text{eq}}$ but when path $B$ contains spin $|-1 \rangle$ it "sees" an equilibrium position of $+\Delta{x}_{\text{eq}}$. If the spin is in state $|0 \rangle$ in either path it "sees" ${x}_{\text{eq}}=0$. This is why, despite at any point in time the difference in forces on each component is identical to the scheme in Ref. \cite{pedernales_2020} for $|-1 \rangle$ and $|0 \rangle$, unlike Ref. \cite{pedernales_2020} both paths deviate from $x=0$, as shown in Fig. \ref{geom}(b). 

\begin{figure*}
	\includegraphics[width=\linewidth]{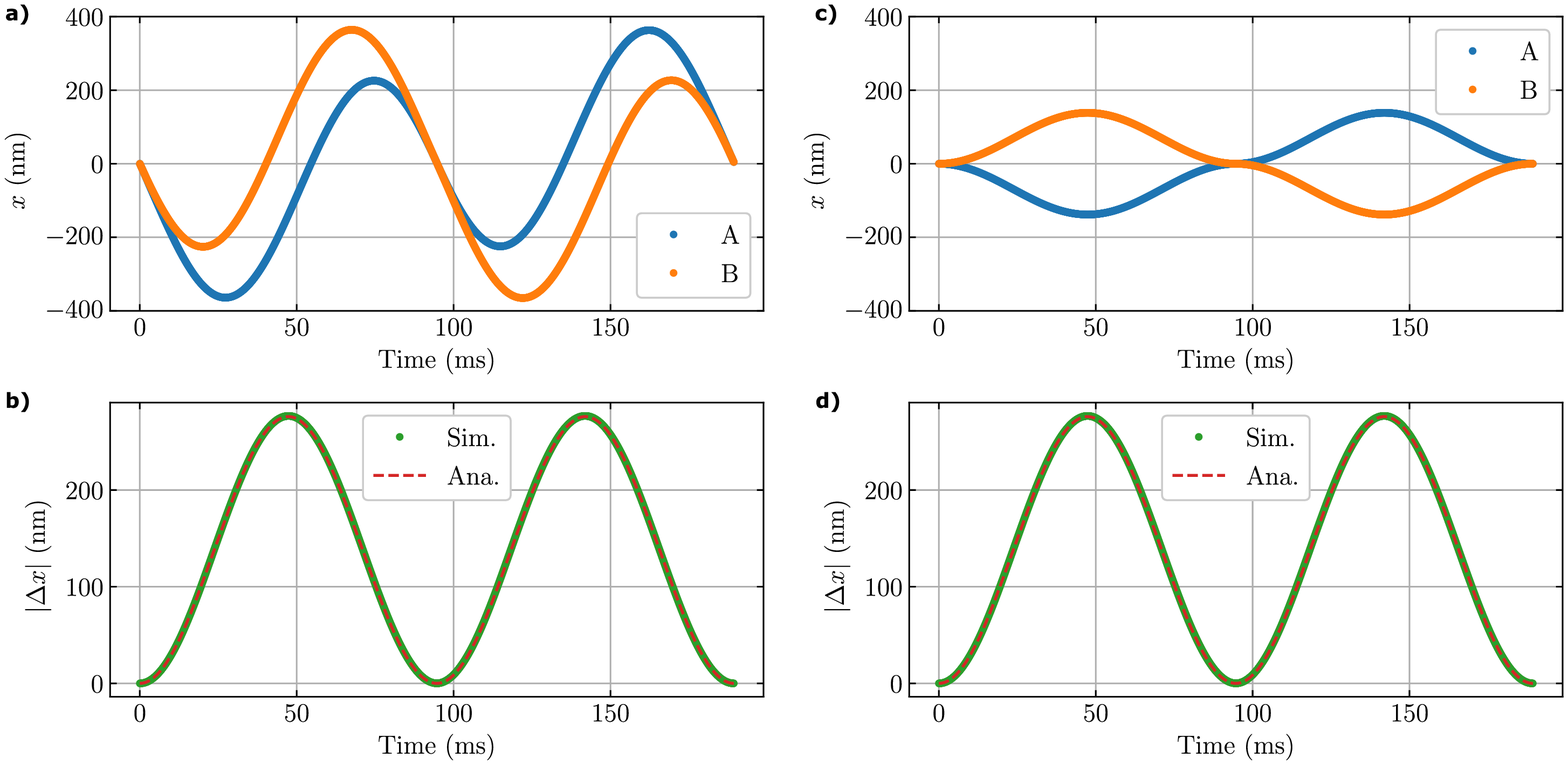}
	\caption{Simulations of the paths taken by the two superposition components in the presence of magnetic teeth, as described by the Hamiltonian of Eq. (\ref{eq:s1}). Along with an extra spin flip at the first recombination of paths, 9800 teeth and the associated spin flips for each tooth crossing are considered with a magnetic field of the form $\mathbf{B}(\hat{\mathbf{r}}) = (\pm{B}^{\prime} \hat{x}+{B}_{0}) {\mathbf{e}}_{x}$, where ${B}^{\prime} = \SI{940}{\tesla\per\metre}$ and ${B}_{0} = \SI{420}{\milli\tesla}$. (a), (b) Paths taken in a geometry where all teeth are the same width. Here, $x$ is the absolute position along the $x$ axis shown in Fig. \ref{geom}. The diamagnetic oscillation of both superposition components is clear. (b) Despite the shared diamagnetic oscillation, the magnitude of the separation between the two components is as described analytically by the Hamiltonian of Eq. (\ref{eq:s3}) \cite{pedernales_2020}. Here, "Sim." labels the simulated separation and "Ana." the analytical. (c), (d) Paths taken in a geometry where the first tooth is half the width of the rest. Here, $|\Delta x|$ is the absolute value of the separation along the $x$ axis between superposition components A and B. The shared diamagnetic oscillation is well canceled, without changing the relative motion of the two components.\label{sim}}
\end{figure*}

As discussed, the neglected ${B}_{0}$ terms have no effect on the relative motion of the superposition components; however, they do induce a diamagnetic oscillation around $x=0$ for both superposition components, as shown in the simulation of the nanodiamond motion for the Hamiltonian of Eq. (\ref{eq:s1}) in Figs. \ref{sim}(a) and \ref{sim}(b). This is not shown in the schematics of the nanodiamond path because it does not change the macroscopicity of the superposition created and, due to the large number of teeth crossed per diamagnetic oscillation, the oscillation can be canceled by having the first tooth be half the width of the rest, as shown in Figs. \ref{sim}(c) and \ref{sim}(d).      

The first three terms of Eq. (\ref{eq:s3}) account for the relative motion of the nanodiamond superposition components as, for small $\phi$, gravity affects the $x$ position of both components identically. That is, for $|\phi | < \SI{1}{\milli\radian}$ and $|\Delta  x| < \SI{300}{\nano\metre}$, it is assumed that at any instant in time both superposition components experience the same $g$ as each other, but that $g$ can change throughout the drop. Therefore, the introduction of tilt only induces a gravitational phase difference between spatially separated superposition components \cite{wan_2016b}. Due to the inversion of paths $A$ and $B$ in the second half of the drop, see Fig. \ref{geom}(b), any gravitational phase difference due to a constant $g$ is canceled \cite{pedernales_2019}. The repeated spin flipping and path inversion also cancels any phase difference due to the zero-field splitting term and both superposition components, $A$ and $B$, contain the spin $|0\rangle$ and $|-1\rangle$ states for the same amount of time. In the ideal case the only phase not canceled between paths $A$ and $B$ is due to the change in $g$ down the $\approx \SI{1}{metre}$ drop as the nanodiamond gets closer to the center of mass of the Earth. This will be discussed further in Appendix \ref{a2}.     

\section{Spin State Interference}\label{a2}

The state of the superposition at time $t$ takes the form \cite{wan_2016b}

\begin{equation}\label{eq:s4}
|\Phi (t) \rangle = \frac{|\psi (t, A) \rangle |A \rangle + {e}^{-i \varphi}|\psi (t, B) \rangle |B \rangle}{\sqrt{2}},
\end{equation}
where the position state of the nanodiamond in path $j$ at time $t$ is given by $|\psi (t, j) \rangle$ and the accumulated phase difference, $\varphi$, due to each spatial component having taken different paths $A$ and $B$ has been explicitly factored out of $|\psi (t, j) \rangle$ for clarity. The spin state $|j \rangle$ of the ${\text{NV}}^{-}$ center in the nanodiamond taking path $j$ is undergoing spin flips at each magnetic tooth crossing:

\begin{equation}\label{eq:s6}
	\begin{gathered}
	|A \rangle = |-1 \rangle\ \rightarrow\ |0 \rangle\ \rightarrow\ |-1 \rangle\ \rightarrow\ |0 \rangle\ \rightarrow\ \ldots\\
	|B \rangle = |0 \rangle\ \rightarrow\ |-1 \rangle\ \rightarrow\ |0 \rangle\ \rightarrow\ |-1 \rangle\ \rightarrow\ \ldots
	\end{gathered}
\end{equation}

Therefore at the instant the spin superposition is created by a $\frac{\pi}{2}$ pulse at time ${t}_{1}$, the state is given by

\begin{equation}\label{eq:s7}
\begin{gathered}
|\psi ({t}_{1}, A)\rangle = |\psi ({t}_{1}, B)\rangle = |\psi ({t}_{1})\rangle, \\
|\Phi ({t}_{1}) \rangle = \frac{|\psi ({t}_{1})\rangle (|A \rangle + |B \rangle)}{\sqrt{2}}.
\end{gathered}
\end{equation}

After the superposition components oscillate once about their respective ${x}_{\text{eq}}$ they recombine at time ${t}_{3}$:

\begin{equation}\label{eq:s8}
\begin{gathered}
|\psi ({t}_{3}, A)\rangle = |\psi ({t}_{3}, B)\rangle = |\psi ({t}_{3})\rangle, \\
|\Phi ({t}_{3}) \rangle = \frac{|\psi ({t}_{3})\rangle (|A \rangle + {e}^{-i{\varphi}_{1}}|B \rangle)}{\sqrt{2}},
\end{gathered}
\end{equation}
where ${\varphi}_{1}$ is the phase difference accumulated by the spatially separated superposition components in an external gravitational field \cite{wan_2016b}. The acceleration due to gravity increases throughout the drop as the nanodiamond moves closer to the center of mass of the Earth. By approximating the acceleration to be constant with the value $g({t}_{2})$, ${\varphi}_{1}$ is given by

\begin{equation}\label{eq:s9}
{\varphi}_{1} = \frac{mg({t}_{2})({t}_{3}-{t}_{1})\Delta {x}_{\text{eq}} \sin(\phi)}{\hbar}.
\end{equation}

The scheme continues for a second oscillation with an extra $\pi$ pulse at ${t}_{3}$, such that the path taken by spin $|A \rangle$ is taken by spin $|B \rangle$ in the second and vice versa, as shown in Fig. \ref{geom}(b). Therefore, at the second recombination at time ${t}_{5}$ the state is given by

\begin{equation}\label{eq:s10}
\begin{gathered}
|\psi ({t}_{5}, A)\rangle = |\psi ({t}_{5}, B)\rangle = |\psi ({t}_{5})\rangle, \\
|\Phi ({t}_{5}) \rangle = \frac{|\psi ({t}_{5})\rangle ({e}^{-i{\varphi}_{2}}|A \rangle + {e}^{-i{\varphi}_{1}}|B \rangle)}{\sqrt{2}},
\end{gathered}
\end{equation}
where ${\varphi}_{2}$ is given by

\begin{equation}\label{eq:s11}
{\varphi}_{2} = \frac{mg({t}_{4})({t}_{5}-{t}_{3})\Delta {x}_{\text{eq}} \sin(\phi)}{\hbar}.
\end{equation}

By discarding a global phase factor the spin state at the point of optical readout is given by

\begin{equation}\label{eq:s12}
\frac{{e}^{-i\Delta\varphi}|A \rangle + |B \rangle}{\sqrt{2}},
\end{equation}
where

\begin{equation}\label{eq:s13}
\begin{gathered}
{t}_{5}-{t}_{3} = {t}_{3}-{t}_{1} = T, \\
\Delta\varphi = \frac{m T\Delta{x}_{\text{eq}}\sin(\phi)(g({t}_{4})-g({t}_{2}))}{\hbar}.
\end{gathered}
\end{equation}

The phase difference of Eq. (\ref{eq:s12}) is observed as an interference pattern in the optical spin readout of the ${\text{NV}}^{-}$ center, after the final $\frac{\pi}{2}$ pulse, as shown in Fig. \ref{geom}(b). The probability that the spin state is measured in $|A \rangle$ is of the form ${P}_{A} = {\cos}^{2}(\frac{\Delta\varphi}{2})$ \cite{wan_2016b}. Therefore, by sweeping the tilt angle $\phi$ the presence of a spatial superposition is evidenced by observing spin interference fringes. For the parameters considered, the fringe width spans a range of approximately $\pm\SI{500}{\micro\radian}$ around $\phi=0$. By having the paths $A$ and $B$ invert in the second oscillation only the change in $g$ contributes to the fringe width. If the scheme ended at ${t}_{3}$, as well as not canceling unwanted relative phase effects due to static potentials \cite{pedernales_2019}, then in Eq. (\ref{eq:s12}) $\Delta\varphi = {\varphi}_{1}$ and the fringe width reduces to $< \SI{1}{\nano\radian}$. Therefore, the two oscillation scheme presented in Figs. \ref{schematic}(b) and \ref{geom}(b) greatly reduces the tilt precision required and cancels unwanted relative phase accumulation.

\bibliography{bib} 

\end{document}